\documentclass[prl,twocolumn,superscriptaddress,showpacs]{revtex4}

\usepackage{graphicx}
\usepackage{longtable}%

\begin{document}

\title{Time-dependent relaxation of strained silicon-on-insulator lines using a partially coherent x-ray nanobeam}%
\author{F.~Mastropietro}%
 \altaffiliation{Present address: IM2NP, Universit\'e Aix-Marseille, France}
 \affiliation{CEA-UJF, INAC, SP2M, 38054 Grenoble, France}
 \affiliation{European Synchrotron Radiation Facility, F-38043 Grenoble, France}
\author{J. Eymery}
 \affiliation{CEA-UJF, INAC, SP2M, 38054 Grenoble, France}
\author{G. Carbone}
 \affiliation{European Synchrotron Radiation Facility, F-38043 Grenoble, France}
\author{S. Baudot}
 \affiliation{CEA-UJF, INAC, SP2M, 38054 Grenoble, France}
 \affiliation{CEA-LETI, Minatec, 38054 Grenoble, France}
\author{F. Andrieu}
 \affiliation{CEA-LETI, Minatec, 38054 Grenoble, France}
 \date{\today}%
\author{V. Favre-Nicolin}
\email{Vincent.Favre-Nicolin@cea.fr}
 \affiliation{CEA-UJF, INAC, SP2M, 38054 Grenoble, France}
 \affiliation{Universit\'e Grenoble-Alpes, Grenoble, France}
 \affiliation{Institut Universitaire de France, Paris, France}

\begin{abstract}
We report on the quantitative determination of the strain map in a strained Silicon-On-Insulator (sSOI) line with a $200\times70\ nm^2$ cross-section. In order to study a single line as a function of time, we used an X-ray nanobeam with relaxed coherence properties as a compromise between beam size, coherence and intensity.  We demonstrate how it is possible to reconstruct the line deformation at the nanoscale, and follow its evolution as the line relaxes under the influence of the X-ray nanobeam.
\end{abstract}


\maketitle

New applications in optoelectronic and electronic semiconductor devices have been achieved by a careful control of strain at the nanoscale level. Several physical properties such as charge carrier mobility in transistors and emission wavelength in quantum dots or well heterostructure have been advantageously improved by applying strain fields adapted to the materials band structure, orientation and doping features \cite{lee_strained_2004, bhattacharya_quantum_2004, pryor_band-edge_2005, jacobsen_strained_2006}.

The measurement of these strain fields has required the development of dedicated techniques with adapted spatial and strain resolution. Electronic imaging techniques have seen tremendous developments and outstanding achievements \citep{hue_direct_2008}, but are always limited by the preparation of thin foil that can considerably relieve internal stress in nanostructures. Very recently, X-ray diffraction has taken profit of the highly brilliant and coherent radiation provided by synchrotron sources \cite{nugent_coherent_2010}. Moreover, the optimization of dedicated focusing optics (compound refractive lenses \cite{snigirev_focusing_1998}, Fresnel Zone Plate (FZP) \cite{jefimovs_fabrication_2007, gorelick_high-efficiency_2011}, Kirkpatrick-Baez mirrors \cite{kirkpatrick_formation_1948, paganin_coherent_2006}) has allowed the use of nanobeams, increasing the spatial resolution of diffraction measurements. This also allowed the use of coherent X-ray diffraction imaging (CXDI) for structure (shape, size) and strain determination of single nano-objects  \cite{ schroer_coherent_2008,robinson_coherent_2009, newton_three-dimensional_2010, newton_phase_2010, jacques_bulk_2011}.

In this letter, we illustrate how the strain of a single strained silicon nanostructure changes during irradiation with x-rays, as a function of measurement time using a partially coherent X-ray nanobeam. Strained Silicon-On-Insulator (sSOI) lines are considered due to their strong interest for enhancing the carrier mobility in metal oxide semiconductors field-effect-transistors (MOSFET) devices \cite{andrieu_impact_2007, baudot_elastic_2009}.

\begin{figure}[!t]
\includegraphics[width=8.5cm]{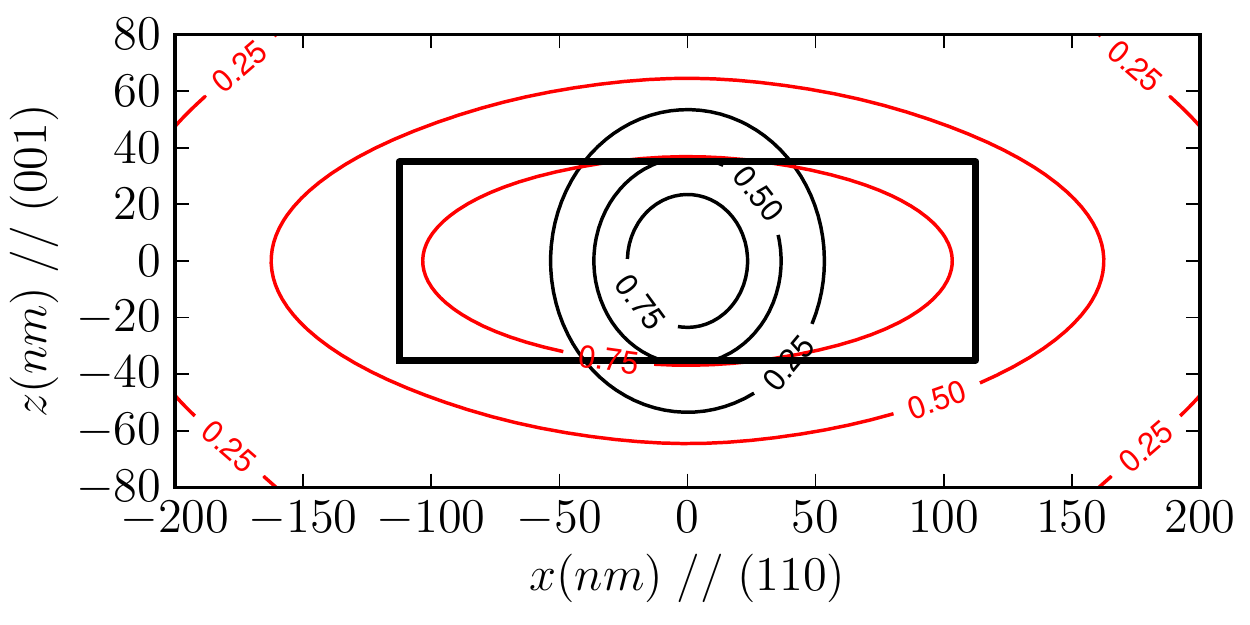}
\caption{Nano-focused X-ray beam at the sample position: (heavy black line) sketch of the rectangular cross-section of the sSOI line, $225\times70\ nm^2$; (black contour levels) normalized intensity distribution at the sample position, calculated using a point source 49 m from the fully illuminated Fresnel Zone plate; (red contour levels) normalized intensity distribution taking into account the source size ($64\times13.8\ \mu m^2$ r.m.s.) for the id01 beamline. Contour levels are represented at 25, 50 and 75 $\%$ of the maximum intensity. The calculations are made at 150 $\mu m$ from the focal point, i.e. the estimated position for the sample. The resulting illumination of the sSOI line is both partially coherent and inhomogeneous. }
\label{figPartialIllum}
\end{figure}

Silicon lines were etched from a $(001)$ oriented sSOI substrate made by a wafer bonding technique from the Si deposition on a SiGe virtual substrate imposing a biaxial strain, as described in \cite{ghyselen_engineering_2004}. Lines in tensile strain ($\epsilon_{yy}=+0.78\%$) are oriented along the $[1\overline{1}0]$ direction which corresponds to the usual direction of n-MOSFET channels for which electron transport is improved. The strain relaxes elastically along $[110]$, i.e. perpendicularly to the lines \cite{baudot_elastic_2009}. An in-plane misorientation of about 1$^o$ is used between the strained Si lines and the Si substrate in order to separate the line and substrate Bragg peaks. The sSOI lines have a width W=225 nm and a height H=70 nm (Fig. \ref{figPartialIllum}) and lie on a 145 nm SiO$_2$ layer. The distance d between two adjacent lines is about 775 nm. Grazing-incidence X-ray diffraction have been performed on these line gratings \citep{baudot_elastic_2009} and gave $\epsilon_{xx}=+0.04\%$ (almost fully relaxed along the line direction [110]) and $\epsilon_{yy}=+0.74\%$, which is very close to the biaxial strain before etching. Linear elasticity calculations allow estimating $\epsilon_{zz}\approx-0.25(5)\%$ from elastic constants.

As recently proven, the displacement field of this system can be probed using coherent beams \cite{newton_phase_2010}. However in the present system the sSOI lines period is about one micron, so that a \textit{single} line can only be studied using a highly focused X-ray beam. The experiment has been conducted on the undulator beamline id01 of the European Synchrotron Radiation Facility, using a 8 keV energy beam obtained with a Si(111) channel cut monochromator. The X-ray beam has been focused to the sample position using a gold FZP with a diameter of 200 $\mu$m and a 70 nm outermost zone width \cite{jefimovs_fabrication_2007}. A beam stop and a pinhole have been used to cut the contribution of the central part of the direct beam and the higher diffraction orders.

The asymmetric ($1\overline{1}3$) Bragg reflection has been probed during the experiment to reduce the contribution of the Si substrate. In addition, in the considered geometry, the detection plane is almost parallel to Ewald's sphere and thus the information about the displacement fields is contained in a single image. The incoming radiation is inclined with respect to the sample surface by $\alpha_i=52.3^\circ$ and the diffracted beam is collected with a Maxipix detector \cite{ponchut_maxipix_2011} at $2\theta_f=56.48^\circ$.

In this experiment the transverse coherence length is estimated to $11(h)\times53(v)\mu m^2$ at the location of the FZP (49 m from the source), due to the combined effects of the undulator source size and the monochromator \cite{diaz_coherence_2010}. Thus, to achieve a highly coherent X-ray beam, a partial illumination of the FZP can be used \cite{mastropietro_coherent_2011}. However, this would lead to a significant increase in the horizontal size of the focused beam (horizontal FWHM of the order of 1 $\mu m$). In order to illuminate a \textit{single} sSOI line, it is therefore necessary to use a partially coherent X-ray beam by using the entire FZP rather than a small area corresponding to the transverse coherence length.

The use of partially coherent X-ray beams has recently been at the focus of a number of studies, \cite{flewett_extracting_2009,whitehead_diffractive_2009} and it was also demonstrated that quantitative reconstructions using \textit{ab initio} phase retrieval algorithms were possible in such a case \cite{clark_high-resolution_2012, huang_three-dimensional_2012}. The present study differs from these previous ones: firstly, the X-ray beam illuminating the FZP is more than 10 times larger than the transverse coherence length in the horizontal direction, thus introducing a significant blur in the diffraction image. Secondly, the size of the focal point ideally produced by a point source placed at 49 m from the FZP is much smaller than the width of the sSOI line. This is illustrated in Fig.\ref{figPartialIllum}, where the cross-section of the sSOI line is compared to the shapes of ideal and partially coherent beams. As the size of the source increases the focal point by a factor $\approx 3$, the simultaneous reconstruction of the object and the coherence function -as proposed in \cite{clark_simultaneous_2011}- becomes difficult.

In order to make a quantitative analysis of the recorded diffraction images, we used a direct approach to calculate the effect of partial coherence and inhomogeneous illumination from the characteristics of the undulator source:
(i) the undulator source was modeled as a superposition of incoherent point sources\footnote{An incoherent description of the source is possible due to the relatively large source size\cite{vartanyants_coherence_2010} used on ID01}, with a Gaussian intensity distribution ($64(h)\times13.8(v)\ \mu m^2$ r.m.s. source size). To limit the computational requirements, a rectangular array of $9\times5$ sources was taken into account.
Then, (ii) the complex amplitude, focused by the FZP, was calculated at the sample position (at each atomic position in a 2D layer of the silicon line), for each of the 45 point sources.
Finally, (iii) the intensity on the detector was calculated as the incoherent sum of the intensities contributed by all the point sources.

\begin{figure}[!t]
\includegraphics[width=8.5cm]{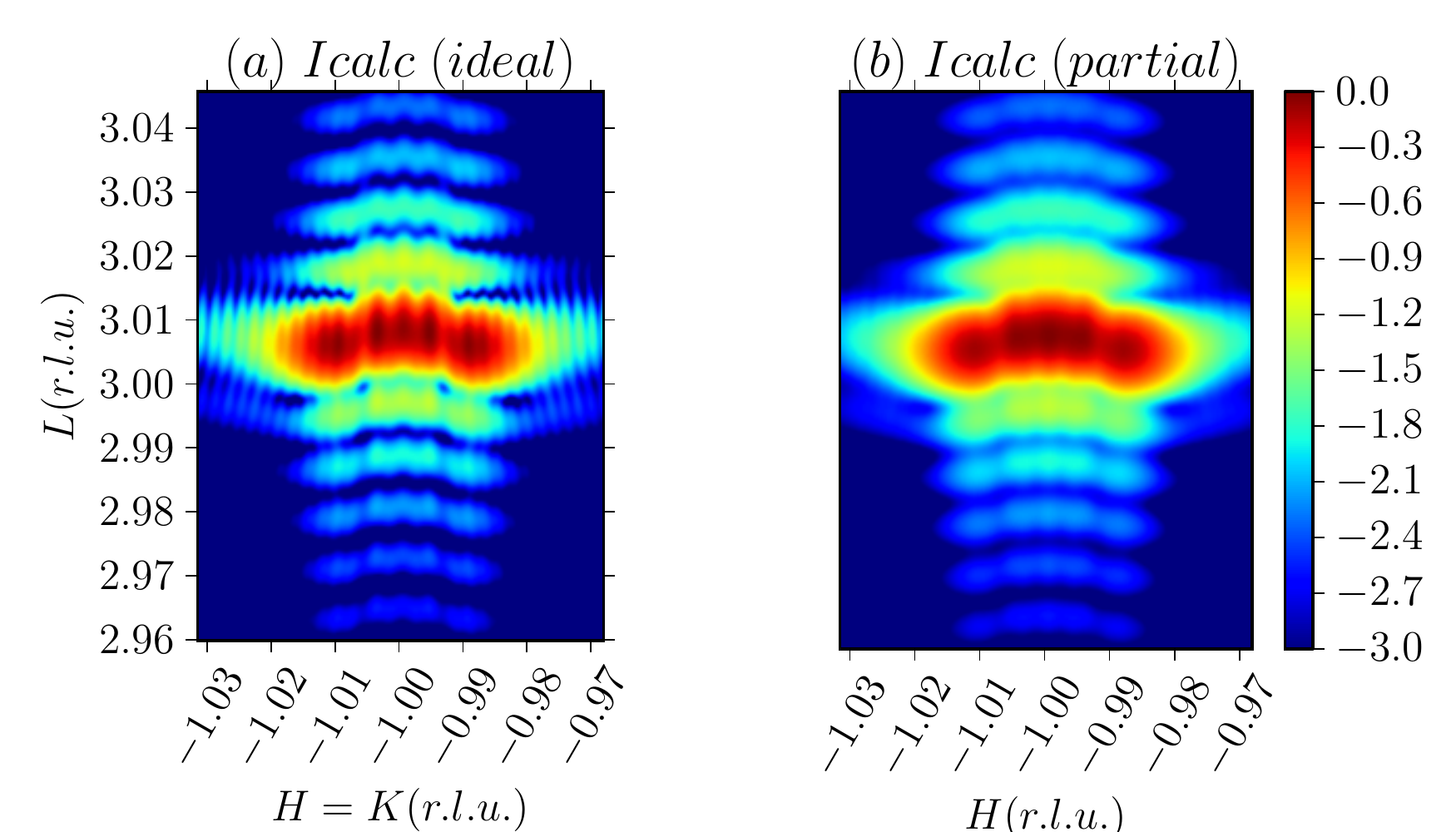}
\caption{Diffraction from a sSOI line by a partially coherent X-ray nanobeam ($(1\overline{1}3)$ reflection), calculated from a simulated deformation of the sSOI line, using either (a) a plane wave illumination or (b) taking into account the partially coherent illumination of the FZP.}
\label{figPartialCohDiff}
\end{figure}

\begin{figure}[!b]
\includegraphics[width=8cm]{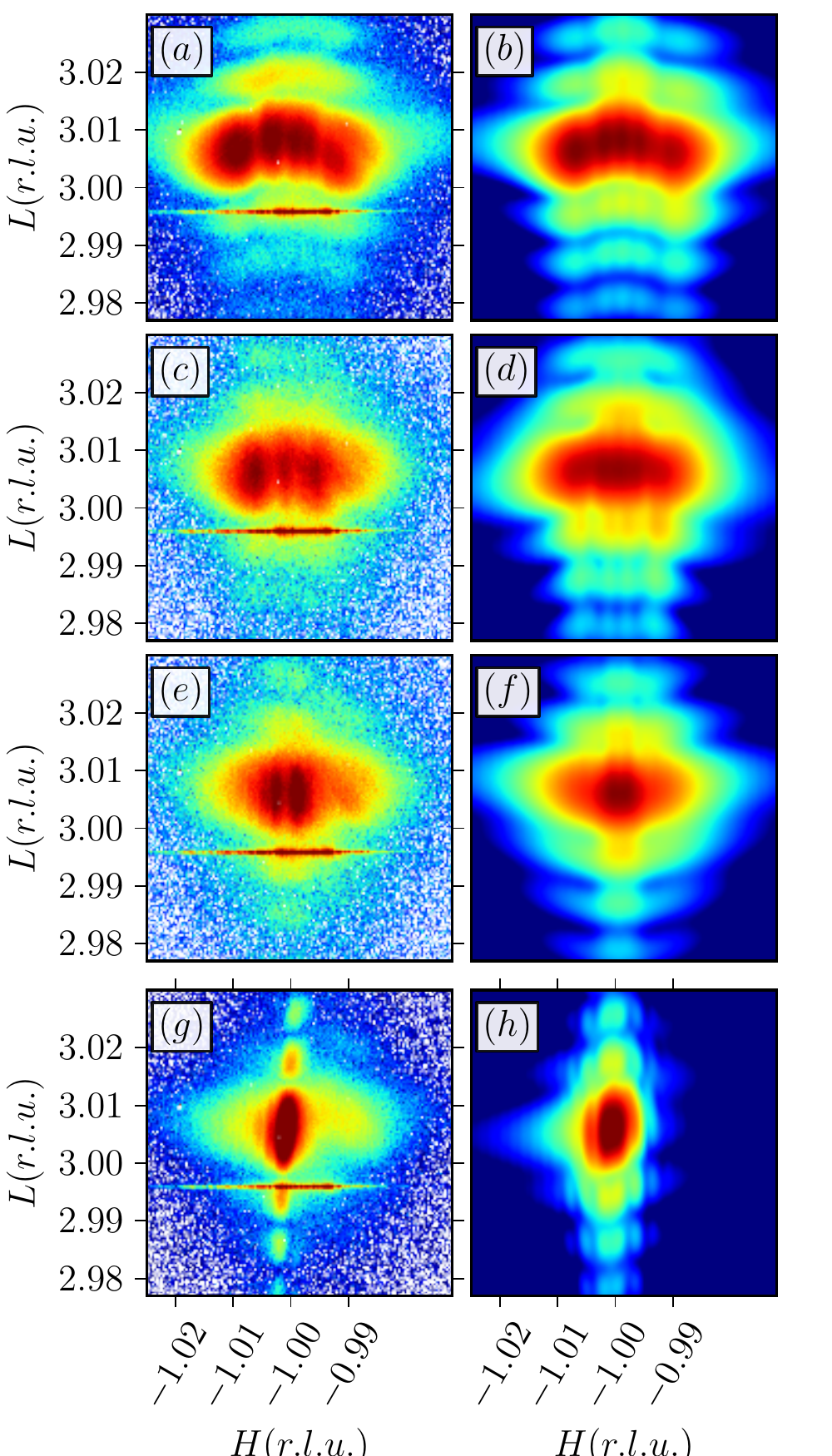}
\caption{(a, c, e, g)Two dimensional diffraction patterns compared to (b, d, f, h) calculations for the chosen silicon line at different times T= 0, 800, 1600 and 2900 s. Relevant changes are visible in the experimental intensity distribution and well reproduced by calculations. Intensities are expressed using the same logarithmic colour scale, with a total amplitude spanning three orders of magnitude. The horizontal stripe near $l=2.996$ is due to the Si substrate, and was masked during the refinement.}
\label{soi_gaussiansource}
\end{figure}

A time-dependent study has been performed illuminating for 3000 s the same section of the selected line with the aim of inducing radiation damage at the Si/SiO$_2$ interface, and investigate its nature and its characteristic time. In our experimental conditions, the x-ray beam carries in term of photon flux $ \approx 5\times10^4$ ph/s/nm$^2$. 
Several 2D diffraction patterns of the described Bragg peak have been collected during the experiment with an acquisition time of 100 s. The most relevant detector images are shown in Fig. \ref{soi_gaussiansource}(a,c,e,g) corresponding to the times: T= 0, 800, 1600, 2900 s.
The oscillations along $L$ are directly related to the thickness ($\approx$ 70 nm) of the SOI line. The 'banana-shaped' intensity distribution observed in the [110] direction  is a consequence of the a spontaneous bending induced by the free-surface strain relaxation already observed \cite{moutanabbir_observation_2011}.  This strain is relaxed during the X-ray exposure. 

In order to determine the deformation field inside the SOI line as a function of time, it was not possible to use standard imaging algorithms based on \textit{ab initio} phase retrieval, due to the high degree of partial coherence. Since the SOI line are perfectly cristalline, we have therefore opted to perform a direct refinement of the 2D displacement field in the plane perpendicular to the direction of the SOI Line. The line was modeled as a perfect silicon crystal, with a $70\times 225\ nm^2$ cross-section, and the displacement of the atoms was described using a polynomial sum: $u_z(x,z)=\sum { a_{n_x,n_z}x^{n_x} z^{n_z}}$, with $0<=n_x<=4$ and $0<=n_z<=2$.

This simple model allows the description of asymmetric displacements inside the SOI line. Diffraction near the $(1\overline{1}3)$ reflection is not sensitive to displacement in the horizontal ($x$ in Fig. \ref{figPartialIllum}) [110] direction (perpendicular to the scattering vector) and the displacement field along the direction of the line ($y$) is assumed to be negligible due to mechanical constraints along this direction.

\begin{figure}[!b]
\includegraphics[width=8cm]{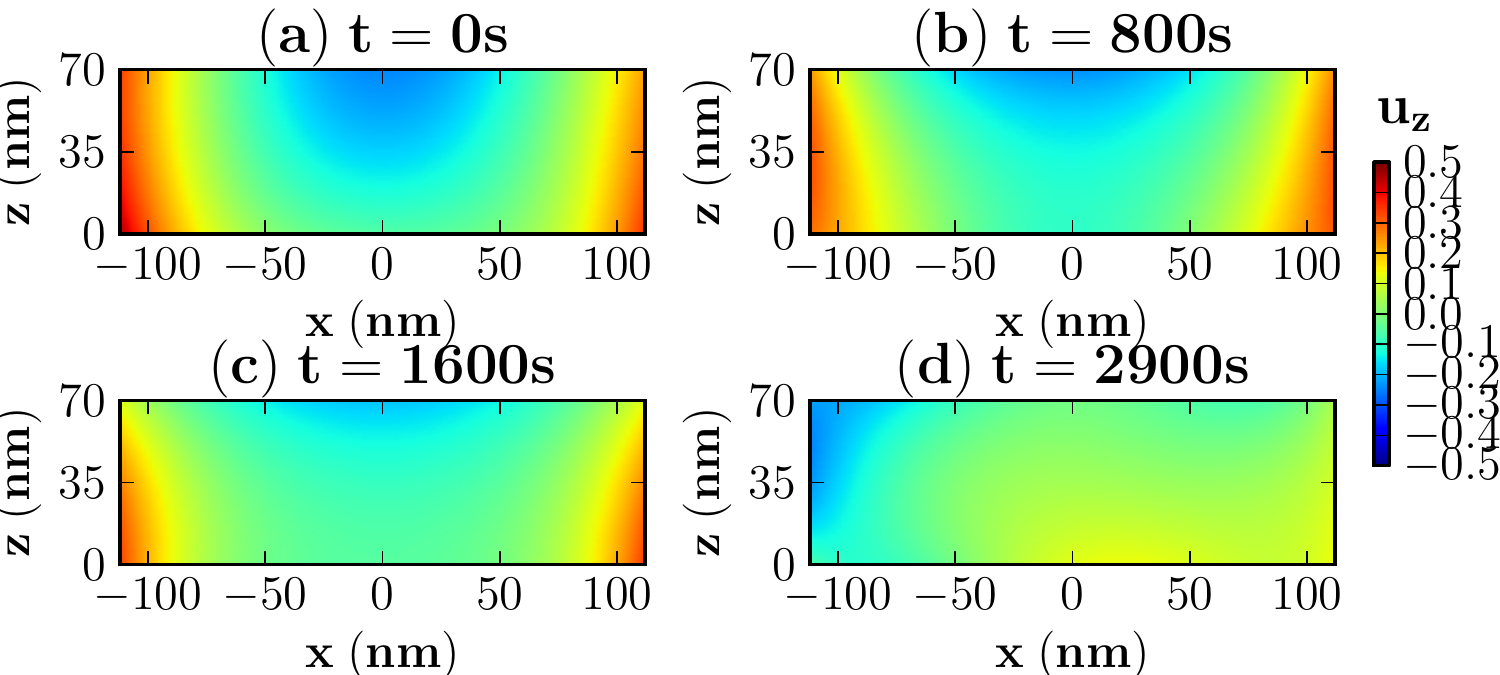}
\caption{Displacement fields $u_z$ obtained from the time-dependent analysis of radiation damage at   T= 0, 800, 1600 and 2900 s. The $u_z$ profiles is visibly less curved at T=2900s (d) with respect to the one obtained at (a) T= 0. Scale units expressed in nanometers are represented by colours.}
\label{uz_profile}
\end{figure}

The scattering is calculated by generating the position of all Si atoms in the $xz$ plane from the $u_z$ displacement field, and uses the PyNX library \cite{favre-nicolin_fast_2011}, which allows fast computing using a graphical processing unit. This method also avoids approximating the scattering as the Fourier transform of the average electronic density multiplied by $exp({2i\pi\vec{s}\cdot \vec{u}})$, which can be incorrect in the presence of a large inhomogeneous displacement field.

The $u_z$ polynomial is minimized using first a parallel simulated annealing algorithm \cite{falcioni_biased_1999}, with 50000 trials, followed by a least squares minimization. The criterion for minimization is $\chi^2=\sum_{i}{\omega_i(I_{obs}^i-\alpha\times I_{calc}^i)^2}$, where $\alpha$ is a scale factor, and the weight $\omega_i$ is equal to $1/I_{obs}^i$ if $I_{obs}^i>0$, and 0 otherwise.

The resulting calculated scattering patterns obtained by fitting the  experimental data are shown in Fig. \ref{soi_gaussiansource}.
The corresponding displacement fields are depicted in Fig \ref{uz_profile}. $u_z$ visibly changes from T=0 s (a) to T=2900 s (d), where it shows a quite less curved profile. The strain fields $\epsilon_{zx}$ and $\epsilon_{zz}$ can be easily calculated from $u_z$: the standard deviations of $\epsilon_{zx}$ and the mean values of $\epsilon_{zz}$ are summarised in Table~\ref{prl:t1}. $<\sqrt{\langle \epsilon_{zx}^2 \rangle}>$ decreases continuously during the relaxation process while $<\epsilon_{zz}>$ remains around the same value. 

\begin{table}[!t]
\caption{Mean strain values at T=0, 800, 1600 and 2900 s.}
  \begin{tabular}{ccccc} \\ \hline
    Strain & 0 s & 800 s & 1600 s & 2900 s \\ \hline
    & & & \\
    $\sqrt{\langle \epsilon_{zx}^2 \rangle}$ & $0.63\%$ & $0.57\%$ & $0.45\%$ & $0.19\%$ \\
    & & & \\
    $\langle \epsilon_{zz} \rangle$ & $-0.19\%$ & $-0.21\%$ & $-0.21\%$ & $-0.18\%$ \\  
    & & & \\ \hline
  \end{tabular}
  \label{prl:t1}
\end{table}

Radiation damage resulting from the highly brilliant beam has been observed both for macromolecular compounds \cite{marchesini_coherent_2003} and semiconductor structure \cite{polvino_synchrotron_2008,favre-nicolin_coherent-diffraction_2009}. A bending of unstrained SOI lines was previously reported and attributed to the underlying oxide structural expansion \citep{shi_radiation-induced_2012}. In this article we show how it is possible to follow quantitatively the evolution of the 2d strain field.

In the present system, the SOI line remains crystalline during the long X-ray exposure, as demonstrated by the continued presence of a Bragg diffraction spot. Moreover, we have shown that the shape of the diffraction peak can be explained by a simple elastic deformation of the line. The relaxation of the line is due to radiation damage occuring at the interface between the silicon line and the underlying SiO$_2$ \cite{polvino_synchrotron_2008}. There is no significant change with time of the period of the fringes corresponding to the 70 nm thickness, indicating that any damage at the interface does not extend significantly into the silicon line.

It should be noted that $<\epsilon_{zz}>$ does not vary despite the clear relaxation : this difference in the evolution between $\epsilon_{zz}$ and $\epsilon_{xz}$ is due to the fact that only the part of the SOI line around the incident beam is relaxed, and therefore the line remains stressed (in tension) where the Si/SiO$_{2}$ interface is not damaged. This is a very important result for MOSEFT, as a significant alteration to the Si/SiO$_{2}$ interface would not affect conduction properties. Finally, the shape of the diffraction spot remains unchanged after waiting ten minutes without illumination from an X-ray beam, indicating that direct heating effects (dilatation) are negligible.

To conclude, we have shown that it is possible to retrieve the strain field inside a single SOI line, and follow its evolution as a function of time during irradiation with an intense X-ray beam. This information can be retrieved even though we used a partially coherent beam in order to optimize both the flux and the beam size on the sample. Moreover, the use of a strongly asymmetric reflection allowed the data collection of a scattering plane containing all the relevant information about the two-dimensional relaxation of the line, using single two-dimensional frames.

The intense irradiation only damaged the Si/SiO$_2$ interface and not the crystalline silicon structure, which furthermore keeps the strong normal strain essential to the conduction properties. This demonstration paves the way for the study of complex, working devices such as metal-oxide-semiconductor field-effect transistor (MOS-FET): this has already been conducted on micron-sized FET, \cite{hrauda_x-ray_2011} but the \textit{in situ} analysis of FET on industrially relevant devices will require very small and intense probes, i.e. X-ray beams exploiting the full intensity of undulator source while avoiding strict coherence constraints.

\begin{acknowledgments}
This work was partially supported by the French ANR XDISPE (ANR-11-JS10-004-01).
\end{acknowledgments}

\bibliography{Article-SOI}

\end{document}